\documentclass[aip,jap,amsmath,amssymb,preprintnumbers,12pt]{revtex4-1}

\usepackage{verbatim}
\usepackage{hyperref}
\usepackage{graphicx}
\usepackage{bm}
\usepackage{sidecap}
\begin{document}

\title{Generation of topologically diverse acoustic vortex beams using a compact metamaterial aperture}%

\author{Christina J. Naify}%
\email{christina.naify@nrl.navy.mil}
\author{Charles  A. Rohde}
\author{Theodore P. Martin}
\author{Michael Nicholas}
\affiliation{U.S. Naval Research Laboratory, Washington, DC 20375, USA , Code 7165}
\date{\today}

\author{Matthew D. Guild}
\affiliation{National Research Council Research Associateship Program, U.S. Naval Research Laboratory, Washington, DC 20375, USA }
\date{\today}

\author{Gregory J. Orris}%
\affiliation{U.S. Naval Research Laboratory, Washington, DC 20375, USA , Code 7165}

\maketitle\clearpage
Vortex waves, which carry orbital angular momentum, have found use in a range of fields from quantum communications to particle manipulation. Due to their widespread influence, significant attention has been paid to the methods by which vortex waves are generated. For example, active phased arrays generate diverse vortex modes at the cost of electronic complexity and power consumption\cite{Hefner1999,Wilson2010,Marzo2015,Riaud2015}. Conversely, analog apertures, such as spiral phase plates\cite{Hefner1999,Wunenburger2015}, metasurfaces\cite{Karimi2014}, and gratings\cite{McMorran2011} require separate apertures to generate each mode. Here we present a new class of metamaterial-based acoustic vortex generators, which are both geometrically and electronically simple, and topologically tunable. Our metamaterial approach generates vortex waves by wrapping an acoustic leaky wave antenna\cite{NaifyLWA2015} back upon itself. Exploiting the antennaÕs frequency-varying refractive index, we demonstrate experimentally and analytically that this analog structure generates both integer, and non-integer vortex modes. The metamaterial design makes the aperture compact and can thus be integrated into high-density systems.

\begin{figure}
\begin{center}
\includegraphics {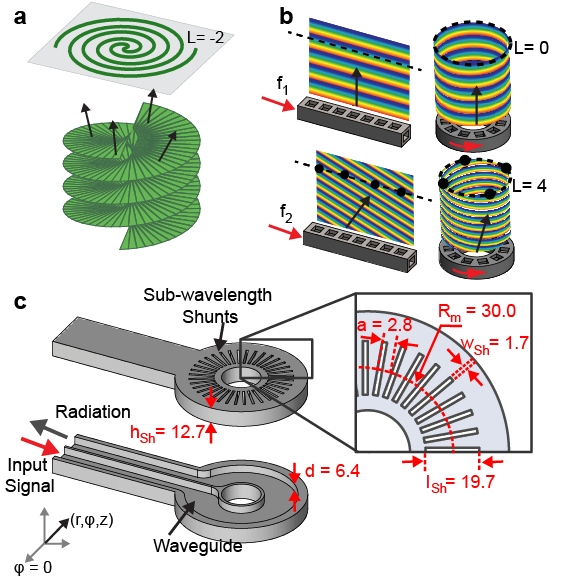}
\caption{Schematic of acoustic vortex wave antenna: (a) Illustration of a propagating vortex wave, and the resulting vortex constant phase contours in a plane parallel to the aperture. The arrows correspond to the wavenumber vectors which have both vertical, and radial components. (b) A one-dimensional leaky wave antenna can be converted to a vortex wave antenna by wrapping the waveguide back on itself. At normal incidence, excited by frequency $f_{1}$, the OAM mode is zero, with no helicity. Increasing the input frequency to $f_{2}$ produces an angled wavefront which wraps to a helical vortex beam. (c) Exploded-view drawing of the acoustic vortex wave antenna (all dimensions in mm). The input signal goes through one side of the waveguide and the radiated port is fully absorbing to prevent reflections. Inset: detail of shunt geometry.}
\label{fig1}
\end{center}
\end{figure}

The total angular momentum of a system can be divided into two components, spin angular momentum, and orbital angular momentum (OAM). Although acoustic waves do not possess spin angular momentum they have been shown to carry OAM~\cite{Hefner1999,Wilson2010}. A drawing of a single mode helical wave with value $L=-2$ is shown in Fig.~\ref{fig1}(a) where $L$ is the OAM topological mode number. A wave with OAM index $L\!=\!0$ describes a system with no helical phase front. The phase front of the propagating wave is a corkscrew-type phase advance with the sign of the topological charge positive, for clockwise rotation, or negative, for counter-clockwise rotation. These vortex waves have been found to be useful in an extremely diverse range of applications from communications\cite{Malr2001,Molina-Terriza2007,Strain2014,Karimi2014,Willner2015,Harris2015} and imaging\cite{Jesacher2005,Bernet2006,Chen2014} to particle manipulation \cite{Grier2003,Ladavac2004,Hong2015}over a wide range of length scales. In the most widely examined application, vortex waves have been harnessed for use in electromagnetic and quantum communications.  

The importance of both topological diversity, and aperture simplicity in vortex mode generation becomes immediately evident when considering the applications of vortex waves. The limitations of existing techniques preclude their use in important applications such as communications, which requires both rapid switching between mode-diverse orthogonal signals, and micro-manipulation, which benefits from compact, simple apertures for large-scale integration. In an effort to address these shortcomings, a recent optical approach utilized scattered whispering gallery modes and demonstrated the ability to generate multiple electromagnetic vortex modes.\cite{Cai2012,Strain2014}. This approach allowed for fast switching of mode value by tuning the refractive index of the emitter to generate a range of whole integer topological values. Here, we show that it is possible, by incorporating sub-wavelength metamaterial elements, to further reduce the relative aperture size while also achieving both integer and fractional vortex modes. Fractional vortex modes further increase vortex wave potential applications as they have been of interest for particle manipulation\cite{Hong2015EPL} and edge-detection imaging\cite{Wang2015sr}.

Inspired by recent research on acoustic leaky-wave antennas\cite{NaifyLWA2013,NaifyLWA2015}, we present an air-acoustic vortex beam emitter which generates topologically diverse vortex waves using a single transducer coupled to a single analog metamaterial aperture. A leaky wave antenna is a device comprised of a one-or two-dimensional waveguide which leaks power along it's length with either a continuous leaking slot or sub-wavelength radiating shunts. Leaky wave antennas rely on geometry-controlled dispersion to tune the refractive index of the fluid inside the waveguide. The leaked energy then refracts from the antenna at an angle $\theta(\omega)$, similar to the refraction mechanism of a prism. The value of $\theta(\omega)$ is determined by the ratio of wavenumber $\beta(\omega)$ inside the waveguide to the wavenumber in the surrounding area $k$\cite{Liu2002}
\begin{equation}
\label{grp}
\theta(\omega)=\operatorname{arcsin}\,(\frac{\beta(\omega)}{k}).
\end{equation}  
Although not included in this study, acoustic leaky wave antennas can be designed with negative refractive indices\cite{NaifyLWA2013}.

Using the simple leaky wave aperture, we have realized a versatile, multi-mode vortex wave antenna by wrapping the linear antenna waveguide back on itself. Consistent with Huygens' Principle, the planar wavefront emitted by a straight leaky waveguide is wrapped into a helical phase front, as illustrated in Fig.~\ref{fig1}(b). By then sweeping through frequency within the waveguide, the wavefront angle, $\theta(\omega)$, changes, resulting in distinct vortex wavefronts. The realized aperture geometry is shown in Fig.~\ref{fig1}(c) and consists of a circular waveguide beneath an array of radially aligned shunts, as well as input and output ports that couple acoustic radiation into the annular region. The resulting continuous, as opposed to discrete, dispersion relation facilitates non-integer mode generation. 

The confinement of the waveguide in the radial ($\hat{r}$) and axial($\hat{z}$) directions imposes constraints that determine the dispersion relation as a function of frequency. In order to solve for the dispersion, it is convenient to use a coordinate transformation \cite{Chin1998}, $u=b\ln(r/b)$ and $v=b\phi$, that unwraps the circular annulus in cylindrical coordinates $(r,\phi,z)$ to form a linear cartesian waveguide with coordinates $(u,v,z)$. After applying the coordinate transform, the wave equation in the plane formed by the annulus becomes,
\begin{equation}
\label{grp}
\frac{\partial^{2}\Psi}{\partial u^{2}}+\frac{\partial^{2}\Psi}{\partial v^{2}}+k^{2}(\omega)\exp(u/b)\left[\exp(u/b)-\frac{i\alpha(\omega)}{k(\omega)b}\right]\Psi=0
\end{equation}  
where $\Psi(u,v)$ is the acoustic pressure in the polar plane, $b$is a scaling parameter of the transform, $k(\omega)=\omega/c$
is the unconstrained wavenumber in air, and $\alpha(\omega)$ depends on the geometry, impedance, and axial confinement of the waveguide. Further details of the analytic model, which takes into account the cumulative end corrections of all shunts, can be found in the supplementary materials. The solutions to Eq.~(\ref{grp}) include a phase term $\exp(iL\phi)$,
where the topological mode $L$ is fixed by the radial and axial boundary conditions and is radially independent,  advancing the phase uniformly at all radii inside the waveguide.  We emphasize that $L$ does not depend on the choice of $b$; the physics of the propagation around the annulus cannot change with the scale of the transformation.  The utility of the scaling parameter $b$ is to determine 
both the dispersion relation $\beta(b,\omega)=L(\omega)/b$ and the radiation angle $\theta(b,\omega)=\arcsin[\beta(b,\omega)/k]$ at a particular radius $r=b$ inside the waveguide.  Traditionally the scaling parameter $b$ has been chosen to represent the centroid of the radial part of the solution $\Psi(r,\phi)$~\cite{Chin1998}; thus one can define a centroid propagation angle based onthis choice.  For a given geometry, $L$ is spatially independent and only depends on frequency; the frequency dependence of $L$ can be tuned by changing the confinement geometry of the waveguide (radial or axial) and/or the shunt geometries.

\begin{figure}
\begin{center}
\includegraphics {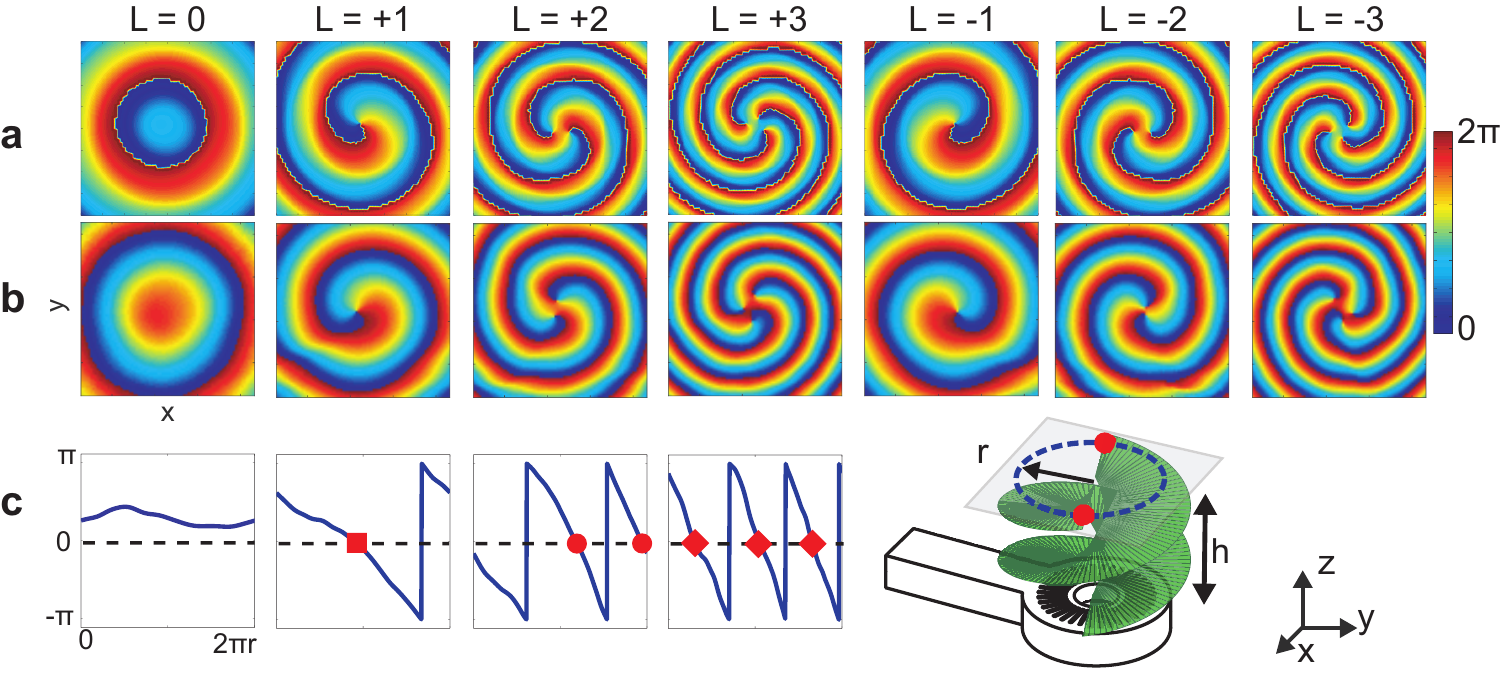}
\caption{Finite element simulated (a) and measured (b) phase distributions of OAM waves with topological charges ranging from -3 to +3: The various charges were generated using a single metamaterial aperture by insonifying the aperture at discrete frequencies. All measurements were taken at a distance of $h$ from the surface of the antenna, where $h$ 50~mm. The evaluated frequncies are 1.8kHz, 2.8kHz, 4.1kHz and 5.45kHz. (c) The method for determination of mode magnitude consists of counting the number of phase fronts which cross a circle that is co-axial with the aperture. The circle, defined on the measurement plane in (a) and (b), is shown in (c, far right). Phase magnitude, collected from the simulated data in (a), is shown with the number of phase fronts indicated by red symbols.}
\label{fig2}
\end{center}
\end{figure}

Using the theory described above, a vortex wave antenna geometry was chosen which is capable of generating at least 7 distinct, orthogonal vortex modes. Finite element methods (FEM) were used to predict the OAM phase topology of each mode, and the design geometry was verified by successful experimental demonstration of the radiated mode. The predicted phase profiles at four frequencies corresponding to whole integer modes are shown in Fig.~\ref{fig2}(a).   

The realized vortex wave antenna was fabricated and characterized.  Figure~\ref{fig2}(b) shows the measured phase topology of 7 modes, extracted from the measured pressure fields, corresponding to $L=-3$ through $L=+3$.  This general design proves extremely versatile. For example, using the same unit cell geometry shown in Fig.~\ref{fig1} and simply changing the aperture's radial midpoint, $R_m$, an entirely new set of mode values are realized. This design flexibility opens the door to vortex beam multiplexing by designing an aperture with multiple concentric waveguides resulting in a multiplexed beam, ideal for high-speed communications applications. Additionally, the versatility of the design space enables tailoring of the aperture radius to be either scaled down even further in size, or increased to minimize vortex beam divergence.

To quantify the momentum mode number of each mode, the phase of the radiated beam has been mapped onto a circle whose axis is concentric with the antenna axis at $r\!=\!0$. The corkscrew wavefront of the radiated beam carries the OAM of the waveguide's dispersion, with phase $\exp(iL\phi)$. Thus for a value of $L$, the phase of a vortex beam will cycle $L$ times through $2\pi$ around the circumference of the axis-concentric circle. Figure~\ref{fig2}(c) shows the phase of the radiated beam on a circle of radius $r\!=\! 0.1$~m. 

\begin{figure}
\begin{center}
\includegraphics {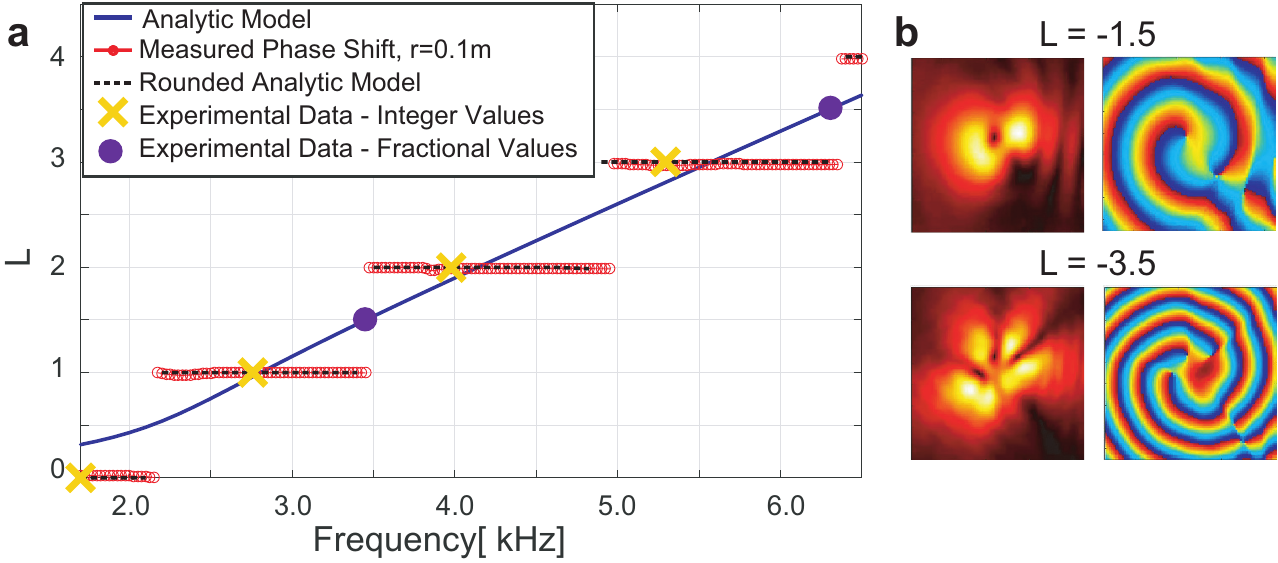}
\caption{Comparison of calculated and measured modes (a), and measured fractional modes (b): (a) Calculated mode value (blue lines). Integer values of L were chosen from the experimental emitted spectrum of the aperture. Measured integer modes are indicated as discrete yellow x-marks, while fractional values are indicated as filled circles. The horizontal lines correspond to (red) the measured mode number determined by phase unwrapping, (black-dashed)  theory rounded to the nearest integer, consistent with mode determination set up by Berry\cite{Berry2004}. (b) Measured pressure (left) and phase (right) amplitudes at frequencies representing fractional vortex values of -1.5 and -3.5.}
\label{fig3}
\end{center}
\end{figure}

The phase at each frequency in the band from 1.8 kHz to 6.3 kHz is unwrapped by adding multiples of $2\pi$ at each phase value of -$\pi$ and integrated along a closed path to determine the total change in phase, $\Delta \Phi$, as a function of frequency. The experimentally obtained values of $\Delta \Phi$ are shown as red circles in Fig.~\ref{fig3}(a) and are observed to discontinuously jump between integer plateaus. This metric indicates that unlike the predicted continuous monotonic change in the topological mode within the waveguide, the mode numbers of the radiated vortex waves are only observed at discrete integer values.  This result, while seemingly paradoxical, is in agreement with previous theoretical and experimental results demonstrated for electromagnetic (EM) vortex waves~\cite{Vasnetsov1998, Berry2004}.  

\begin{figure}
\begin{center}
\includegraphics {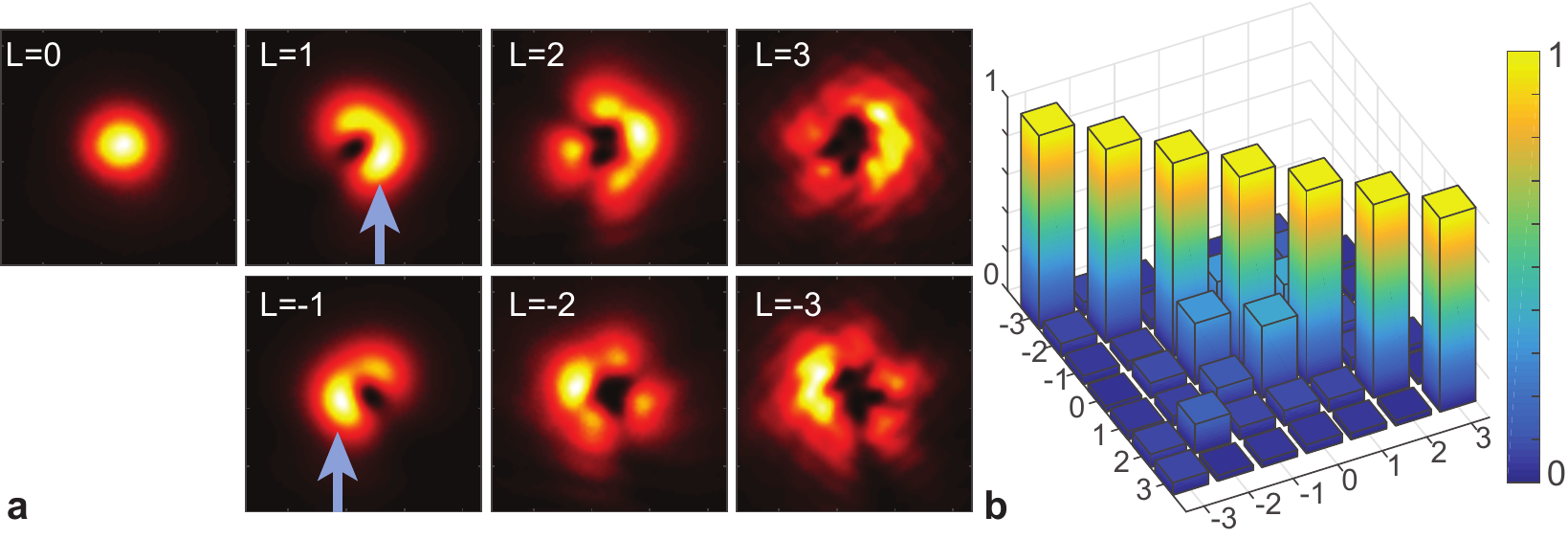}
\caption{Measured mode intensities and calculated orthogonality: (a) Measured acoustic intensity plots of $L=0$, $\pm1$, $\pm2$, and $\pm3$ modes measured respectively at frequencies 1.8kHz, 2.8kHz, 4.1kHz and 5.45kHz. Arrow indicates input port direction for the top and bottom rows. (b) Symmetric mode orthogonality matrix for all $L$ modes shown in (a). }
\label{fig4}
\end{center}
\end{figure}

Fractional ordering of the topological modes can instead be observed in the vorticity structure of the radiated beam~\cite{Berry2004}. Integer modes have no phase discontinuity around the annulus (see Fig.~\ref{fig2}(b))and are expected to produce a single radiated vortex, with an accompanying null in amplitude (see Fig.~\ref{fig4}(a)), that is centered on the beam's axis. Fractional modes, on the other hand, result in multiple observed vortices, with associated nulls in amplitude, centered around the beam axis~\cite{Berry2004}. The number of vortices arising from a fractional topology is given by the closest integer to $L$, and are predicted to increase discontinuously when $L\!=\!N+1/2$, where $N$ is an integer.  

As the phase approaches a half-mode, a radially-oriented line of vortex pairs (and associated nulls) appears in the radiated beam profile~\cite{Berry2004,Vasnetsov1998}. This pattern can be observed in the measured phase and amplitude profiles as a dislocation in the wavefront, shown in Fig.~\ref{fig3}(b) for $L \!=\! -1.5$ and $L \!=\! -3.5$ at frequencies where the unwrapped phase is observed to transition between integer modes. The introduction of these fractional modes can be seen in the supplementary material. Since these vortex pairs are equal in magnitude but have opposite signs, summation along a closed path leads to no net contribution to the total vorticity. As the value of $L$ increases beyond a half integer, one of these line vortices breaks off to to form the next higher integer vortex singularity in the cluster, while the others disappear~\cite{Berry2004}.  It has been theoretically predicted that in free space this should lead to a rapid jump from one integer mode to the next, which is in excellent agreement with the observed experimental results obtained using the unwrapped-phase metric presented in Fig.~\ref{fig3}(a). This fractional effect, in which a dislocation in the wavefront appears, has also been described as an analog to the Aharonov-Bohm effect with a flux line consisting of wavefronts along a nodal surface.\cite{Berry2004,Berry1980}

Inspection of the amplitude nulls in Fig.~\ref{fig4}(a) indicates that multiple vortex singularities are also present at the frequencies which predict integer values for the cases of $L=\pm2$ and $\pm3$. Finite element simulation indicates that this apparent fractional topology is a result of the aperture design, where the input and output ports cause the design to be missing a wedge compared to a perfect circle. The missing wedge acts as an effective phase discontinuity at the aperture, which produces a beam profile that appears to result from a fractional phase dispersion. With an ideal design in which the input and output ports are removed and a continuity condition placed on the wrapped waveguide, perfect integer modes would be achievable. We emphasize that the ideal vortex wave antenna concept (with continuity condition) is capable of generating not only whole integer modes, but fractional vortex modes as well (observed in Fig.~\ref{fig3}(b)), which have found a myriad of uses in their own right.  

Since mode orthogonality is vital for many applications of vortex waves including high-speed communications and imaging, the covariance matrix of the measured modes was calculated. Figure~\ref{fig4}(a) shows the measured intensity ($\| p \|^2$) distributions for the $L=0$, $\pm1$, $\pm2$, and $\pm3$ modes. The covariance matrix of the measured modes is plotted in Fig.~\ref{fig4}(b). The details of the calculation can be found in the supplementary materials. The covariance matrix shows that despite the lack of symmetry in the  mode shapes, each mode distribution is highly uncorrelated with the others, indicating a high degree of orthogonality. 

Our acoustic vortex wave antenna approach offers an extremely compact method to emit vortex modes with the diameter of the aperture being the size of the wavelength of excitation at the highest frequency examined. The small, electronically and geometrically simple device makes this vortex wave antenna ideal for both scaling, and for integration into systems requiring a large number of vortex beams. Acoustic OAM waves have already shown usefulness in micromanipulation, such as an acoustic tractor beam, and for fluid mixing. Although these concepts can be also achieved using optical vortex beams, acoustic particle manipulation has advantages over optical manipulation in that it gives the ability to manipulate both larger objects and non-magnetic or non-conducting materials\cite{Ding2012,Mulvana2013}. 

\section*{Acknowledgements}
Work is supported by the Office of Naval Research.

\section*{Author Contributions}
C.N. conceived the idea and designed the aperture. C.R. and M.N. collected the experimental data and performed the data analysis. T. M. and M. G. derived the analytical transformation calculation. G.O. provided guidance on the research direction and the applications. All authors contributed to writing the paper.

\section*{Competing Financial Interests}
The authors declare no competing financial interests.

\section*{Supplementary Material}

\section{Materials and Methods}

Finite element simulations (FEM) of the vortex wave were performed using COMSOL multiphysics. The FEM used a circular waveguide with geometry indicated. The boundary conditions on the waveguide were rigid in the radial direction and a continuous radiation condition was imposed on the exit port of the waveguide. In order to be consistent with the geometry realized experimentally, a wedge of the waveguide was removed corresponding to the input and output ports in the fabricated system. The circular waveguide aperture is fed via the input port using an omnidirectional source. This was simulated in the finite element model using a source signal in the form of a continuous wave at constant amplitude. Perfect absorbing conditions were mandated at the exit port of the waveguide to prevent reflection. Mode chirality was tuned by switching the location of the source between input and output ports.

The leaky wave aperture was fabricated using additive manufacturing in two pieces as indicated in Figure 1. The dimensions were chosen to provide an antenna which would have the best performance at the frequencies for which the transducers had optimal performance. Soft putty was used to seal the top and bottom plates and prevent sound leakage. The aperture design incorporates an input/output waveguide. The omnidirectional sound source was coupled into this waveguide using a 1~m long pipe in order to mitigate the effects of reflection. The output port was filled with sound-absorbing foam. 
The source profile was an linear frequency modulated (LFM, or chirped) pulse. The LFM pulse was 6ms in duration and windowed with a Hanning envelope function. The LFM pulse was a linear ramp of frequencies from 1kHz to 10kHz (generated with the MATLAB \emph{chirp} and \emph{hann} functions). This waveform was digitized with a Agilent 33220A waveform generator and fed to a Dayton Audio CE32A-8 speaker. Efforts were made to mitigate reflection in the radiation area by using anechoic foam. A microphone (Bre\"ul and Kj\ae r 4939-A) mounted on an automated two-dimensional positioning system was used to map the emitted intensity and phase profile above the aperture. Multiple 10ms long waveforms were collected and averaged together to fully capture the 6ms LFM, accounting for waveguide dispersion and collection time of flight. As in the simulated result, chirality was varied by switching input and output ports. In all predicted and measured phase and intensity plots the aperture is centered on in the image. The phase and intensity profiles are mapped at a plane parallel to the surface of the aperture over an area of 0.16~m$^2$ square and at a height of 0.05~m from the top of the aperture.

\section{Vortex Leaky Wave Antenna: Analytic Model}

\subsection{Introduction}

The circular acoustic leaky wave antenna (LWA) is designed
using a ring antenna architecture in cylindrical coordinates. Our 
design assumes propagation around an annular ring in the azimuthal
$\hat{\phi}$ direction, with waveguide confinement in the radial
$\hat{r}$ and axial $\hat{z}$ directions. There are quasi-radiating
boundaries at the input/output ports of the ring (Fig.~1(C) of the main text) that allow an acoustic wave to
adiabatically enter and leave the circular waveguide with minimal reflection
(to first order). Therefore, for purposes of an analytic approximation we assume no
confinement in the azimuthal direction. The wave equation in cylindrical
coordinates is,
\begin{gather}
\label{eqn:1} \frac{1}{r}\frac{\partial}{\partial r}\left(r\frac{\partial P}{\partial r}\right)+\frac{1}{r^{2}}\frac{\partial^{2}P}{\partial\phi^{2}}+\frac{\partial^{2}P}{\partial z^{2}}+k^{2}P=0
\end{gather}
where $P(r,\phi,z)$ is the acoustic pressure and $k=\omega/c_{0}$ is the wavenumber
in air. We assume an acoustic wave of the form $\tilde{P}\exp(i\omega t-i\vec{k}\vec{\cdot x})$
such that a negative imaginary wavenumber results in loss. The annular
waveguide has acoustic shunts in its upper axial plane that allow
the acoustic wave to radiate in the positive $\hat{z}$ direction.
The shunts have rectangular cross-section but are aligned along the
radial direction and have a constant azimuthal spacing $\phi_{a}$.
The dimensions of the acoustic shunts and waveguide are given in Table~I.

\begin{table}[b]
\label{tab:1}
\centering
\begin{tabular}{|c|c|c|}
\hline 
Parameter & Value (mm) & Description\tabularnewline
\hline 
\hline 
$l_{Sh}$ & 19.7 & Shunt length in $\hat{r}$ direction\tabularnewline
\hline 
$w_{Sh}$ & 1.7 & Shunt width in $\hat{\phi}$ direction\tabularnewline
\hline 
$h_{Sh}$ & 12.7 & Shunt depth in $\hat{z}$ direction\tabularnewline
\hline 
$R_{i}$ & 17.3 & Inner radius of Shunt/Waveguide\tabularnewline
\hline 
$R_{o}$ & 42.7 & Outer radius of Shunt/Waveguide\tabularnewline
\hline 
$R_{m}$ & 30.0 & Midpoint radius of Shunt/Waveguide\tabularnewline
\hline 
$d$ & 6.5 & Waveguide height in $\hat{z}$ direction\tabularnewline
\hline 
$a$ & 6.7 & Shunt spacing at $R_{o}$ in $\hat{\phi}$
direction\tabularnewline
\hline 
\end{tabular}

\caption{Dimensions of the leaky wave antenna.}

\end{table}

The waveguide's cross-sectional dimensions are smaller than the acoustic
wavelength $\lambda\simeq43$ mm at the highest experimental frequency
of 8 kHz. Therefore we expect a single propagating mode, confined
in the $\hat{r}$ and $\hat{z}$ directions, that traverses the circular LWA in the $\hat{\phi}$ direction.
The LWA will radiate a vortex wave that is dependent on the topological charge $L(\omega)$, which 
indexes the orbital angular momentum of the waveguide's azimuthal mode.
One must solve Eq.~\ref{eqn:1} under the confinement constraints of the waveguide in
order to estimate the frequency dependence of the topological charge.

\subsection{Solutions to the Cartesian Coordinate Transform}

To solve for the dispersion relation of the annular propagation, we employ a coordinate transformation that
has been successfully applied to electromagnetic ring resonators~\cite{Chin1998}, 
$u=b\ln(r/b)$ and $v=b\phi$, with $z$ unchanged. Here $b$ is a scaling parameter that has units of radius.  The
transformation maps the cylindrical annulus to a linear cartesian
waveguide with coordinates $\left(u,v,z\right)$. Under this coordinate
transformation the wave equation becomes,
\begin{gather}
\label{eqn:2} \frac{\partial^{2}P}{\partial u^{2}}+\frac{\partial^{2}P}{\partial v^{2}}+\exp(2u/b)\frac{\partial^{2}P}{\partial z^{2}}+k^{2}\exp(2u/b)P=0.
\end{gather}
Using separation of variables $P(u,v,z)=F(u,z)G(v)$ we assign a plane
wave solution $G(v)=C\exp(i\beta v)+D\exp(-i\beta v)$ to describe the propagation along the (transformed) linear waveguide, where the propagation constant $\beta$ defines the radiation angle in the transformed system.
The wave equation now becomes, 
\begin{gather}
\label{eqn:3} \frac{\partial^{2}F}{\partial u^{2}}+\exp(2u/b)\frac{\partial^{2}F}{\partial z^{2}}+\left[k^{2}\exp(2u/b)-\beta^{2}\right]F=0.
\end{gather}
An exact solution $F(u,z)$ to Eq.~\ref{eqn:3} is not separable due to the radial dependence of our shunt geometry; however, one can exploit the subwavelength scale of the axial confinement to derive an approximate solution that is separable.  We assume that $F(u,z)=\psi(u)\chi(z)$ with $\chi(z)=B\cos\left(\gamma z\right)$ under the condition that $\gamma d\ll 1$.  Note that $d$ is the maximal value of $z$ in the waveguide and the rigid lower boundary is placed at $z=0$.  Under this assumption the confinement parameter $\gamma$ can be written as,
\begin{gather}
\label{eqn:4} \gamma \simeq \sqrt{\frac{ikw_{Sh}}{\phi_{a}d}\frac{Z_{0}}{Z_{Sh}(\omega)}\frac{1}{b\exp(u/b)}},
\end{gather}
which can be derived from the impedance boundary condition of the shunts at the axial upper boundary ($z=d$). Here $Z_{Sh}(\omega)$ is the acoustic impedance of an individual shunt and $Z_{0}$ is the impedance of air.  Although $\gamma$ depends explicitly on $u$, it is easy to show that the non-separable terms in Eq.~\ref{eqn:3} are proportional to $(\gamma z)^{2}$ and are therefore negligible in the limit $\gamma d\ll 1$.  In this limit the wave equation can be written as,
\begin{gather}
\label{eqn:5} \frac{\partial^{2}\psi}{\partial u^{2}}+\left\{ k^{2}\exp(u/b)\left[\exp(u/b)-\frac{i\alpha(\omega)}{kb} \right]-\beta^{2}\right\} \psi=0, \\
\label{eqn:6} \alpha(\omega)=\frac{w_{Sh}}{\phi_{a}d}\frac{Z_{0}}{Z_{Sh}(\omega)}.
\end{gather}

After applying rigid boundary conditions at $u_{i}(R_{i})$ and $u_{o}(R_{o})$, and transforming back to cylindrical coordinates, the solution to Eq.~\ref{eqn:5} in the plane of the ring is,
\begin{gather}
\label{eqn:7} \Psi(r,\phi)=\psi(r)G(\phi)=A\, r^{L} \exp[\pm iL\phi-ikr] \left( \Omega[\mu,2L+1,2ikr]+\xi \Lambda[-\mu,2L,2ikr] \right), \\
\label{eqn:8} \mu=\left(2L+1+\alpha \right)/2,
\end{gather}
where $\Omega[\mu,2L+1,2ikr]$ is the confluent hypergeometric
function, $\Lambda[-\mu,2L,2ikr]=\Lambda_{-\mu}^{2L}(2ikr)$ is the generalized
Laguerre polynomial, and $L=\beta b$ is the topological charge.  The parameter $\xi$ defines the proportionality between $\Omega[...]$ and $\Lambda[...]$, is $L$-dependent but spatially independent, and is derived from the radial boundary conditions. The topological charge $L$ is invariant under the choice of $b$, and is spatially independent for a given waveguide geometry.  $L$ indexes the azimuthal phase in the same manner as the Bessel order in the solution to the unconfined cylindrical wave equation, except that here $L$ can take on non-integer values.  

\subsection{Impedance of the shunts}

The acoustic impedance of each shunt is determined from the dynamic
mass response of its internal air column, viscous end corrections
at the openings, and mass end corrections due to radiation from the
shunt and its neighbors. The internal mass response is proportional
to the shunt depth $h_{Sh}$~\cite{Chocano2015},
\begin{gather}
\label{eqn:9} Z_{h}=i\omega\rho_{0}h_{Sh}\left[1-\frac{\tanh(s \sqrt{i})}{s\sqrt{i}}\right]^{-1} \\
\label{eqn:10} s=w_{Sh}\sqrt{\frac{\omega\rho_{0}}{\eta_{0}}}
\end{gather}
where $\eta_{0}$ is the dynamic viscosity of air. The viscous and
mass end corrections are discussed at length by Ingard~\cite{Ingard1953}.
The viscous end correction can be approximated as~\cite{Chocano2015,Ingard1953},
\begin{gather}
\label{eqn:11} Z_{v}=\sqrt{8\eta_{0}\omega\rho_{0}}.
\end{gather}
The mass end corrections $Z_{e,i}$ can be calculated using a double
integration over the cross-sectional area of the shunt and its neighbors~\cite{Temkin2001},
\begin{gather}
\label{eqn:12} Z_{e,i}\simeq \frac{i\omega \rho_{0}}{2\pi l_{Sh}w_{Sh}} \int_{A} \int_{A'} \frac{e^{ikR'}}{R'} dA' dA
\end{gather}
where $Z_{e}$ and $Z_{i}$ are impedance corrections at the 
external and internal ends of the shunt, respectively.  Here $dA$ is an infinitesimal 
area element of the shunt's cross-section, $dA'$ is a cross-sectional area element of the shunt or its neighboring shunts,
and $R'$ is the displacement between elements $dA$ and $dA'$~\cite{Temkin2001}. The total shunt impedance is then the series
combination of each component, $Z_{Sh}=Z_{h}+Z_{v}+Z_{e}+Z_{i}$. 
 
The double integral in Eq.~\ref{eqn:12} is intractable for a rectangular cross-section, and series
approximations have been carried out by Ingard~\cite{Ingard1953}. We use numerical integration 
to estimate $Z_{e,i}$, where the integration is carried out over all 35 shunts for the external end correction 
while the internal correction includes only the nearest and second-nearest neighboring shunts. The integral in Eq.~\ref{eqn:12} 
is multiplied by a phasing function $\exp(i\Phi)$ for each neighbor that approximates the phase change around the ring, 
where the value of $\Phi$ is estimated from the measured topological charge at a given frequency. Additionally, mirror reflections from the rigid boundaries inside the waveguide will result in virtual neighbors; we include the first set of these virtual neighbors (one above $R_{o}$ and one below $R_{i}$) in the estimate of the internal end correction. 

As the curvature of the waveguide grows, direct displacement vectors between shunts inside the waveguide will become disrupted by the waveguide's walls, resulting in displacement vectors involving multiple reflections. Given that the displacement $R'$ for reflected vectors is highly sensitive to the wall curvature, and that the 
phase $\Phi$ is not constant around the ring, we assume that the contributions of multiple internal reflections from distant neighbors will largely cancel out. Therefore, there is some amount of error in choosing the number of shunts to include in the calculation of the internal end correction. The near field corrections due to neighboring shunts become stronger at lower frequencies, approaching a magnitude similar to $Z_{h}$ at frequencies below $L=1$; therefore the uncertainty is strongest at these frequencies.

\section{Calculation of Spatial Mode Orthogonality}

The following procedure was used to compute the spatial mode covariance matrix. This is a measure of the spatial mode orthogonality. The norm of each mode, $N_m$, was computed using the Fourier transformed pressure time series data, $p({\bf{x}}, t)$\cite{Tao2014}:

\begin{equation}
\int_\Omega P_m({\bf {x}}, f) P_m^*({\bf {x}}, f) d\Omega \equiv N_m
\end{equation}
where $P_m$ is the Fourier transform of the collected time series data, the mode number $m  = -3,-2, -1, 0, +1, +2, +3$ was selected at discrete frequencies, $f$, guided by the spiral wave condition $m= 2 \pi \beta r$ of Fig. 3 in the main text. $\Omega$ is the sample plane of the mode distribution, and $P^*$ indicates complex conjugation. The covariance between two separate modes, $m$, and $n$ is given by: 
\begin{equation}
C_{mn} = \int_\Omega \frac{1}{\sqrt{N_m}}P_m({\bf {x}}, f) \frac{1}{\sqrt{N_n}}P^*_n({\bf {x}}, f) d\Omega
\end{equation}
and were approximated with the discrete formulation:
\begin{equation}
\textrm{C}_{mn} \simeq \frac{1}{M}\sum_{j=1}^M \frac{1}{\sqrt{N_m}}P_m({{x}_j}, f) \frac{1}{\sqrt{N_n}}P^*_n({{x}_j}, f) 
\end{equation}
with $j$ indexing each position in the collected sample planes. A discrete Fourier transform was used to calculate the complex frequency domain pressure amplitudes and phases. Sufficient padding was added to the 7ms, Tukey windowed time series data sets to create a $\Delta f=25$Hz. The full results of the pressure and phase evolution as a function of frequency can be found in the animations included in the supplementary materials.

\end{document}